\documentclass[]{ceurart}

\usepackage{listings}\usepackage{hyperref}
\usepackage{tikz}
\usepackage[dvipsnames]{xcolor}

\definecolor{colorblind1}{RGB}{1,115,178}
\definecolor{colorblind2}{RGB}{222,143,5}

\makeatletter
\renewcommand{\@copyrightLine}{%
  \tiny
  \textcopyright\ \@copyrightyear\ \@copyrightclause
}
\makeatother

\ExplSyntaxOn
\RenewDocumentCommand{\xmpdatawrite}{m m}
  {
    \str_case:nnF {#1}
      {
        {CopyrightURL} { }
        {Journaltitle} { }
        {Journalnumber} { }
      }
      {
        \xmpdata_write:n
          {
            \iow_char:N \\ #1
            \iow_char:N \{ #2 \iow_char:N \}
          }
      }
  }
\ExplSyntaxOff

\begin{document}

\copyrightyear{2026}
\copyrightclause{Wyatt Olson, Jared Soundy, and J. Z. Schroeder.
  All rights reserved.}

\conference{Presented at the Pragmatics of SAT (PoS) Workshop, FLoC 2026, July 19, 2026, Lisbon, Portugal.}

\title{Exploring 3-CNF Formulas Generated From Partial Steiner Triple Systems}

\author{Wyatt Olson}[%
email=Wyatt.Olson@trojans.dsu.edu,
]
\address{Dakota State University,
  820 N Washington Ave., Madison, South Dakota 57042, United States of America}

\author{Jared Soundy}[%
orcid=0000-0003-2354-0598,
email=Jared.T.Soundy@dsu.edu,
]
\cormark[1]

\author{J. Z. Schroeder}[%
orcid=0000-0002-7303-9120,
email=Justin.Schroeder@dsu.edu,
]
\cormark[1]

\cortext[1]{Corresponding author.}

\begin{abstract}
Steiner triple systems (STS) and subsets of these known as partial STS (PSTS) are combinatorial objects that have received substantial interest from mathematicians since their introduction in 1847.
Over the last three decades, (P)STS have found applications in diverse fields ranging from digital circuit testing to wireless transmission.
Perhaps unsurprisingly, the triple in (P)STS allows for a mapping to 3-conjunctive normal form (3-CNF) formulas, the Boolean formula used as input to 3-SAT.
Prompted by this ready conversion of (P)STS to 3-CNF formulas and numerous real-world applications of (P)STS, we explore the intersection of (P)STS with 3-SAT and its counting version \#3-SAT.
We explain the properties of (P)STS-generated 3-CNF formulas in terms familiar to the SAT community and experimentally compare them against randomly-generated 3-CNF formulas.
We find that PSTS-3-CNF formulas are statistically distinct from randomly generated 3-CNF formulas, in both probability of satisfiability and expected solution count.
\end{abstract}


\maketitle

\section{Introduction}

Cross-fertilization between academic disciplines has repeatedly proven fruitful: the use of category theory structures in programming practice \cite{MOGGI199155}, algebraic geometry and number theory in cryptography \cite{koblitz1987elliptic}, and convex geometry in approximation algorithms for combinatorial optimization \cite{goemans1995improved}, just to name a few.
In this paper, we explore the intersection of combinatorial designs known as \textit{Steiner triple systems} (STS) with 3-SAT and \#3-SAT.

\begin{figure}
	\centering
    \begin{tikzpicture}[every node/.style={draw,circle,fill=black,inner sep=2pt,minimum size=8pt},scale=1.25]

        \begin{scope}[xshift=0mm]
    		\node (0) at (1,1.73) {};
    		\node (1) at (.5,0.865)  {};
    		\node (2) at (0,0) {};
    		\node (3) at (1,0.577)  {};
    		\node (4) at (1,0)  {};
    		\node (5) at (1.5,0.865)  {};
    		\node (6) at (2,0)  {};
    		
    		\draw[line width=1pt] (0)--(2)--(6)--(0)--(4) (1)--(6) (2)--(5);
    		\draw[dashed,line width=1pt] (3) circle (0.577);

            \node[draw=none,fill=none] at (1,-1) {STS(7)};
        \end{scope}

        \begin{scope}[xshift=.25\textwidth]
    		\node (a0) at (0,1.73) {};
    		\node (a1) at (0,0.865)  {};
    		\node (a2) at (0,0) {};
    		\node (b0) at (0.865,1.73) {};
    		\node (b1) at (0.865,0.865)  {};
    		\node (b2) at (0.865,0) {};
    		\node (c0) at (1.73,1.73) {};
    		\node (c1) at (1.73,0.865)  {};
    		\node (c2) at (1.73,0) {};
    		
    		\draw[line width=1pt] (a0)--(a1)--(a2) (b0)--(b1)--(b2) (c0)--(c1)--(c2);
            \draw[line width=1pt] (a0)--(b0)--(c0) (a1)--(b1)--(c1) (a2)--(b2)--(c2);
            \draw[line width=1pt] (a0)--(b1)--(c2) (a2)--(b1)--(c0);

            \draw[loosely dashed,line width=1pt] (a1)--(b0) to[out=45,in=45,looseness=2.35] (c2);
            \draw[densely dashdotted,line width=1pt] (b0)--(c1) to[out=315,in=315,looseness=2.35] (a2);
            \draw[loosely dashdotted,line width=1pt] (c1)--(b2) to[out=225,in=225,looseness=2.35] (a0);
            \draw[densely dashed,line width=1pt] (b2)--(a1) to[out=135,in=135,looseness=2.35] (c0);

            \node[draw=none,fill=none] at (0.865,-1) {STS(9)};
        \end{scope}
		
	\end{tikzpicture}
	\caption{The unique STS(7) and STS(9), with blocks given by the solid straight lines and the dashed circle or arcs.
    Although there exist exactly one STS(7) and STS(9), there is a combinatorial explosion in the number of distinct $\text{STS}(n)$s as $n$ grows.}
    \label{fig:sts_examples}
\end{figure}
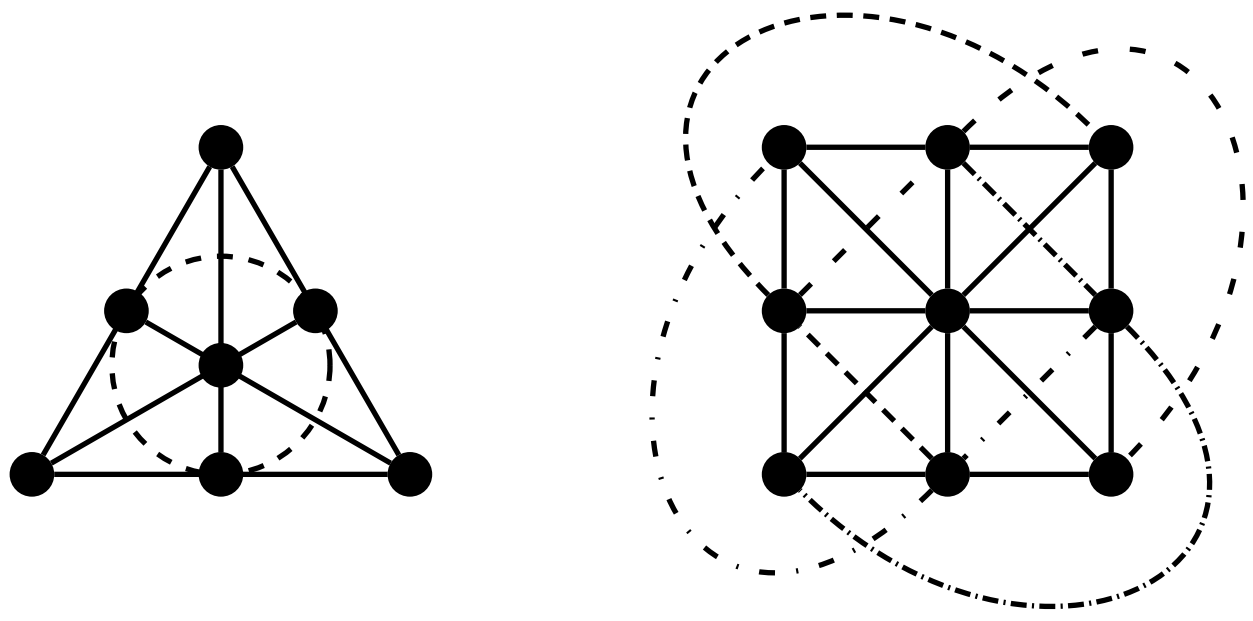

An STS($n$) is a set of $n$ elements together with a set of unordered triples such that each pair of unique elements appears in exactly one triple (\textit{e.g.}, Figure \ref{fig:sts_examples}).
The existence of these structures was initially established by Kirkman \cite{kirkman1847stsexist} in response to a more general problem posed by Woolhouse \cite{woolhouse1844firststs}. They are perhaps best known for their connection to Kirkman's schoolgirl problem \cite{kirkman1850schoolgirl}, in which he asked if it were possible to send 15 students on daily walks in groups of three such that students would walk with every other student exactly once over the course of a week.
STS have proven useful in several areas beyond mathematics:
multiple-input multiple-output space shift keying in wireless transmission \cite{eltira2024steiner,mesleh2022novel, mesleh2022massive},
disk-arrays \cite{830330, COHEN2003103, moura2006distributed},
digital circuit testing \cite{fujiwara2010combinatorial},
watermarking \cite{yang2014watermark},
and cryptography \cite{zhao2024combinatorial}.
Several of these studies \cite{COHEN2003103, moura2006distributed, fujiwara2010combinatorial, zhao2024combinatorial} are noteworthy because in addition to STS they also consider partial STS (PSTS), where the ``exactly one triple'' requirement is relaxed to ``at most one triple'' (\textit{e.g.}, Figure \ref{fig:psts_examples}). 

If (P)STS have proven useful in so many fields, might they not find application in the 3-SAT and \#3-SAT communities?
A natural place to begin is using (P)STS for 3-CNF formula generation\footnote{Note that this is exactly opposite of the limited existing work considering STS and 3-SAT \cite{hartley1996steiner, OSK2025strongstarters}.}
and comparing them against an existing generation method.
Briefly put, we take the points of the PSTS to be variables and map the triples to clauses.
For example, in Figure \ref{fig:psts_examples} every point is a variable, every line is a clause, and literals are set to positive or negative randomly.
One aspect that makes PSTS interesting under the lens of satisfiability and counting problems is the way in which their variables are distributed.
In particular, these PSTS-3-CNF problems have a co-occurrence of at most one between any two variables\footnote{By contrast a full STS-3-CNF formula as a co-occurrence of \emph{exactly} one between any two variables.}.
Additionally, the number of STS($n$)s and PSTS($n$)s grows extremely rapidly---there are over 14 quadrillion STS(21)s, for example \cite{heinlein2023enumerating}---making them a vast space from which to create unique (P)STS-3-CNF problems.
Since a full STS($n$) exists if and only if $n\equiv 1,3$ (mod 6) and has a fixed number of triples $m$, using PSTS provides more flexibility, particularly regarding the parameters $n$, $m$, and---consequently---the ratio $m/n$.

\subsection{Contributions:}
\begin{enumerate}
    \item We introduce a novel method to generate 3-CNF formulas from PSTS.
    \item We generate more than 1.5 million PSTS-3-CNF and uniform random 3-CNF formulas, and compute the number of satisfying assignments for each formula.
    \item We complete a statistical analysis to compute the probability of satisfiability and the expected number of satisfying assignments given satisfiability.
    \item We show with statistical significance that PSTS-3-CNF are more likely to be satisfiable than uniform random 3-CNF formulas and differ in the expected number of satisfying assignments.
\end{enumerate}

\section{Related Work} \label{sec:related}

A Steiner triple system (STS) consists of a set $V$ of points and a set $\mathcal{B}$ of unordered triples such that every pair of points from $V$ is contained in exactly one triple from $\mathcal{B}$. This generic definition allows ample flexibility in how to interpret both the triples and the underlying points. Initial interest in STS was driven by recreational pursuits like Kirkman's schoolgirl problem \cite{kirkman1850schoolgirl}, in which the points represented a class of students broken up into groups of three. Modern applications use the points and triples of an STS to represent, for example, the check disks and information disks of a RAID architecture \cite{830330,COHEN2003103,moura2006distributed}, the bits and codewords of a response compaction circuit known as an X-code \cite{fujiwara2010combinatorial}, the rows and columns of a digital watermark represented as a matrix \cite{yang2014watermark}, the transmit antennas and subsets of active antennas in MIMO wireless communication networks \cite{eltira2024steiner,mesleh2022novel,mesleh2022massive}, or the shares and sets of distributed shares in a repairable ramp scheme \cite{zhao2024combinatorial}. 

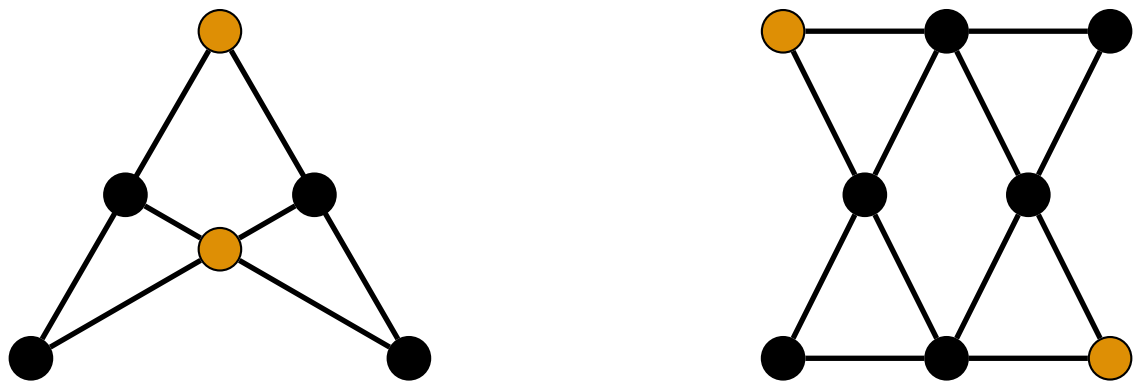
\begin{figure}
    \centering
    \begin{tikzpicture}[every node/.style={draw,circle,fill=black,inner sep=2pt,minimum size=8pt},scale=1.25]

        \begin{scope}[xshift=0mm]
    		\node[fill=colorblind2] (0) at (1,1.73) {};
    		\node (1) at (.5,0.865)  {};
    		\node (2) at (0,0) {};
    		\node[fill=colorblind2] (3) at (1,0.577)  {};
    		\node (5) at (1.5,0.865)  {};
    		\node (6) at (2,0)  {};
    		
    		\draw[line width=1pt] (0)--(2)--(3)--(5) (1)--(3)--(6)--(0);

            \node[draw=none,fill=none] at (1,-0.45) {PSTS(6)};
        \end{scope}

        \begin{scope}[xshift=.25\textwidth]
    		\node[fill=colorblind2] (c0) at (0,1.73) {};
    		\node (b01) at (0.4325,0.865)  {};
    		\node (a0) at (0,0) {};
    		\node (c1) at (0.865,1.73) {};
    		\node (b12) at (1.2975,0.865)  {};
    		\node (a1) at (0.865,0) {};
    		\node (c2) at (1.73,1.73) {};
    		\node[fill=colorblind2] (a2) at (1.73,0) {};
    		
    		\draw[line width=1pt] (a0)--(a1)--(a2) (c0)--(c1)--(c2);
            \draw[line width=1pt] (a0)--(b01)--(c1) (a1)--(b12)--(c2);
            \draw[line width=1pt] (c0)--(b01)--(a1) (c1)--(b12)--(a2);

            \node[draw=none,fill=none] at (0.865,-0.45) {PSTS(8)};
        \end{scope}
    \end{tikzpicture}
    \caption{
    A PSTS allows every pair of points to share a triple at most once, whereas an STS requires they share a triple exactly once.
    Notice the orange points in each diagram above do not share a triple; hence, these are PSTS, not STS.
    }
    \label{fig:psts_examples}
\end{figure}

If we replace the word ``exactly'' with ``at most'' in the definition of STS, we arrive at a partial STS (PSTS); see Figures \ref{fig:sts_examples} and \ref{fig:psts_examples} for examples of STS and PSTS, respectively. 
Theoretical research on PSTS has included topics like embedding PSTS in full STS, an NP-complete problem \cite{colbourn1983embedding}, and classifying STS by the sub-PSTS they do or do not contain \cite{ganschow2026antiperfect,heinlein2022algorithms,wolfe2005sparse}.
Note that \cite{830330,COHEN2003103,moura2006distributed,fujiwara2010combinatorial,zhao2024combinatorial} all involve STS that avoid having any sub-PSTS(6) isomorphic to the Pasch configuration shown in Figure \ref{fig:psts_examples}.
In short, PSTS have a rich theoretical literature, in addition to the real-world applications they have found in these earlier cited papers: \cite{COHEN2003103, moura2006distributed, fujiwara2010combinatorial, zhao2024combinatorial}.

Connections between combinatorial designs and satisfiability problems have also been explored previously. For example, Hartley \cite{hartley1996steiner} constructed SAT instances whose satisfying assignments would yield small Steiner systems (some of which were STS), and Ogandzhanyants \textit{et al.} \cite{OSK2025strongstarters} used satisfying assignments to a specific SAT instance in a tripling construction for strong starters. Both of these studies used generally hard-to-solve SAT problems to create combinatorial objects; this paper aims to do the opposite: create SAT problems from existing combinatorial objects.

There exists a wide variety of sources for 3-CNF formulas for both 3-SAT and \#3-SAT, from real-world problems \cite{biere1999symbolic} to contrived problems \cite{hartley1996steiner, cheeseman1991really, mitchell1992hard}.
Families of problems differ in their characteristics in measurable ways beside construction \cite{mitchell1992hard}.
The benchmarks that represent these diverse families frequently prove valuable in both solver competition and real-world application settings \cite{xu2008satzilla}.
Further exploration and publication of random and structured families prevent our assessment of solvers from missing important blind spots similar to those noted in \cite{mitchell1992hard}.

It is well known that the clause-to-variable ratio has a high impact on hardness and probability of satisfiability \cite{cheeseman1991really}.
Therefore, our statistical analysis considers the ratio as an independent variable, and we generate 3-CNF formulas for a range of ratios.
This range includes the well-known phase transition point of 4.26 \cite{cheeseman1991really, nudelman2004understanding}.

It is observed in \cite{mitchell1992hard} that hard problems are not necessarily representative of common problems in practice, and that they are often contrived to illustrate the worst cases; however, in order to test SAT solving algorithms in a way that measures their robustness, these cases must be considered.

\section{Definitions} \label{sec:def}

We use standard definitions of 3-CNF formulas. For any positive integer $n$, let $[n]=\{1,2,...,n\}$.
Let there be a set of $n$ binary variables $V = \{x_1,x_2,..., x_n\}$ and a set of $m$ clauses $\Sigma = \{C_1,C_2,..., C_m\}$, where for each $j\in[m]$ the clause $C_j = (\ell_{j} \lor \ell_{j^\prime} \lor \ell_{j^{\prime\prime}})$ is the disjunction of exactly three literals.
Let there be corresponding literals for every variable $x_i \in V$ such that, for each $j \in [m]$ and $t\in\{j,j^\prime,j^{\prime\prime}\}$, the literal $\ell_{t} \in \{x_i, \neg x_i\}$ for some $i \in [n]$.
The 3-CNF formula is then the conjunction of all such clauses in $\Sigma$.

Formally, a {\em partial Steiner triple system of order $n$}, briefly PSTS($n$), is an ordered pair $(V,\mathcal{B})$\ where $V$ is a set of $n$ elements (often called {\em points}) and $\mathcal{B}$ is a set of 3-element subsets of $V$ called {\em triples} such that every pair of distinct points from $V$ is contained in at most one triple from $\mathcal{B}$ (see Figure \ref{fig:psts_examples}).
In a slight abuse of notation, we take the point set $V=\{x_1,x_2,...,x_n\}$ to be a set of binary variables.
If every pair of distinct points from $V$ is contained in \emph{exactly} one triple from $\mathcal{B}$ (opposed to \emph{at most} one), then we can drop the word ``partial'' and refer to the ordered pair as a {\em Steiner triple system of order $n$}, or STS($n$); compare the STS in Figure \ref{fig:sts_examples} with the PSTS in Figure \ref{fig:psts_examples}. While a PSTS($n$) can exist for any positive integer $n$, it is well known that an STS($n$) exists if and only if $n\equiv 1,3$ (mod 6) \cite{kirkman1847stsexist}. An STS($n$) will have $n(n-1)/6$ triples, so the ratio of triples to points is fixed at $(n-1)/6$.
For example, an STS(31), STS(49), or STS(63) will have a triple-to-point ratio of 5, 8, or 10.33, respectively.
Using our (P)STS to 3-SAT mapping (introduced below), this will result in clause-to-variable ratios of 5, 8, or 10.33.
This has clear implications on the probability of satisfiability for the resulting 3-SAT problems and motivates our use of PSTS, which do not have a ratio solely dependent on $n$.

To formally define our conversion from PSTS to 3-SAT, we need to define a mapping from triples to clauses.
To allow for randomly negating variables, let $S$ be a Bernoulli random variable representing the sign of a literal.
For each occurrence of a point $x_i \in V$ in some triple $b \in \mathcal{B}$, we will draw an independent realization of $S$, denoted $s_{x_i,b}$, where $P(s_{x_i,b}=1)=0.5$ and $P(s_{x_i,b}=0)=0.5$.
Let $\mathcal{C}$ be the set of all possible clauses.
We define the function $g:\mathcal{B}\rightarrow \mathcal{C}$ by $g(\{x_i,x_{i^\prime},x_{i^{\prime\prime}}\})=\ell_i\lor \ell_{i^\prime}\lor \ell_{i^{\prime\prime}}$ where for each $t\in\{i,i^\prime,i^{\prime\prime}\}$:

\[
\ell_t=
\begin{cases}
    x_t, & \text{if } s_{x_t,b}=1\\
    \neg x_t, & \text{if } s_{x_t,b}=0.
\end{cases}
\]

\noindent Essentially, each clause is formed from a triple in the PSTS by randomly negating each variable with a probability of 0.5. The resulting set of clauses $\Sigma=g(\mathcal{B})$ forms a 3-CNF formula on the set of variables $V$. 
For brevity, we refer to CNF formulas generated by this method as PSTS-3-CNF.

We use the typical definitions of the problems 3-SAT and \#3-SAT: determining whether a 3-CNF formula is satisfiable and how many satisfying assignments it contains, respectively.
We compare and contrast PSTS-3-CNF with uniformly random generated 3-CNF formulas (U-3-CNF) \cite{nudelman2004understanding}, rejecting tautology and duplicate clauses.

\section{Experiment}

Here we describe our data generation methods for PSTS-3-CNF and U-3-CNF across a range of variable counts, $n$, and clause-to-variable ratios, $r = \frac{m}{n}$.
Values of $n$ are chosen from 35 to 65 in increments of 5.
The upper range of 65 was determined by computational limits associated with our large sample size.
As noted in the related work, it is well known that ratio is an important factor in 3-SAT, therefore we account for it as an independent variable and choose the following ratios for data generation: $\{3.0, 3.2, \cdots, 5.0\}$.
On the low-end, this range captures PSTS-3-CNF and U-3-CNF that are nearly always satisfiable, to nearly always unsatisfiable on the high-end.
For each $n$ and desired ratio $x \in \{3.0, 3.2, \cdots, 5.0\}$, we set $m=\lfloor n\cdot x\rfloor$ so that $r = \frac{m}{n} \approx x$.
Combined, these choices result in seven unique values of $n$, eleven ratios $r$, and two generation methods (PSTS-3-CNF and U-3-CNF), thereby creating 154 combinations of independent variables.
For each combination of variables, we generate 10000 formulas, totaling 1.54 million formulas.

Our PSTS generation method relies on adapting Revised Stinson's Algorithm described by \cite{Stinson1999}.\footnote{
The Revised Stinson's Algorithm was implemented in a previous project by Joshua Ganschow under the mentorship of the third author. His source code can be found at \href{https://github.com/GanschowJosh/Steiner-Triple-Systems-Research}{https://github.com/GanschowJosh/Steiner-Triple-Systems-Research}.}
Revised Stinson's Algorithm builds a PSTS one triple at a time by taking a random point $x$ and finding two additional random points $y$ and $z$ that have not yet been used in a triple with $x$; the triple $\{x,y,z\}$ is added to the PSTS, and if $\{y,z\}$ were previously contained in a triple with a third point different from $x$, that triple is removed. Thus, at each step of the algorithm, the number of triples in the PSTS will increase by one or stay the same, and we simply repeat this process until the desired number of triples has been reached.
From there the PSTS is converted to 3-CNF as described in Section \ref{sec:def}.
Our U-3-CNF generation is built from methods described in \cite{mitchell1992hard}.

We use the probabilistic exact model counter Ganak to determine the number of satisfying assignments for all CNF formulas \cite{sharma2019ganak, soos2025ganak}.
We acknowledge that Ganak is not an exact model counter; however, its theoretical guaranteed error rate is incredibly low (based on the probability of a hash collision) \cite{sharma2019ganak, soos2025ganak}, and empirical validation showed no errors on benchmarks \cite{sharma2019ganak}.
Ganak's GitHub page lists error rate as ``...often less than 1 case per billion....'' \cite{ganak_github}

\begin{figure}
    \centering
     \includegraphics[width=\textwidth]{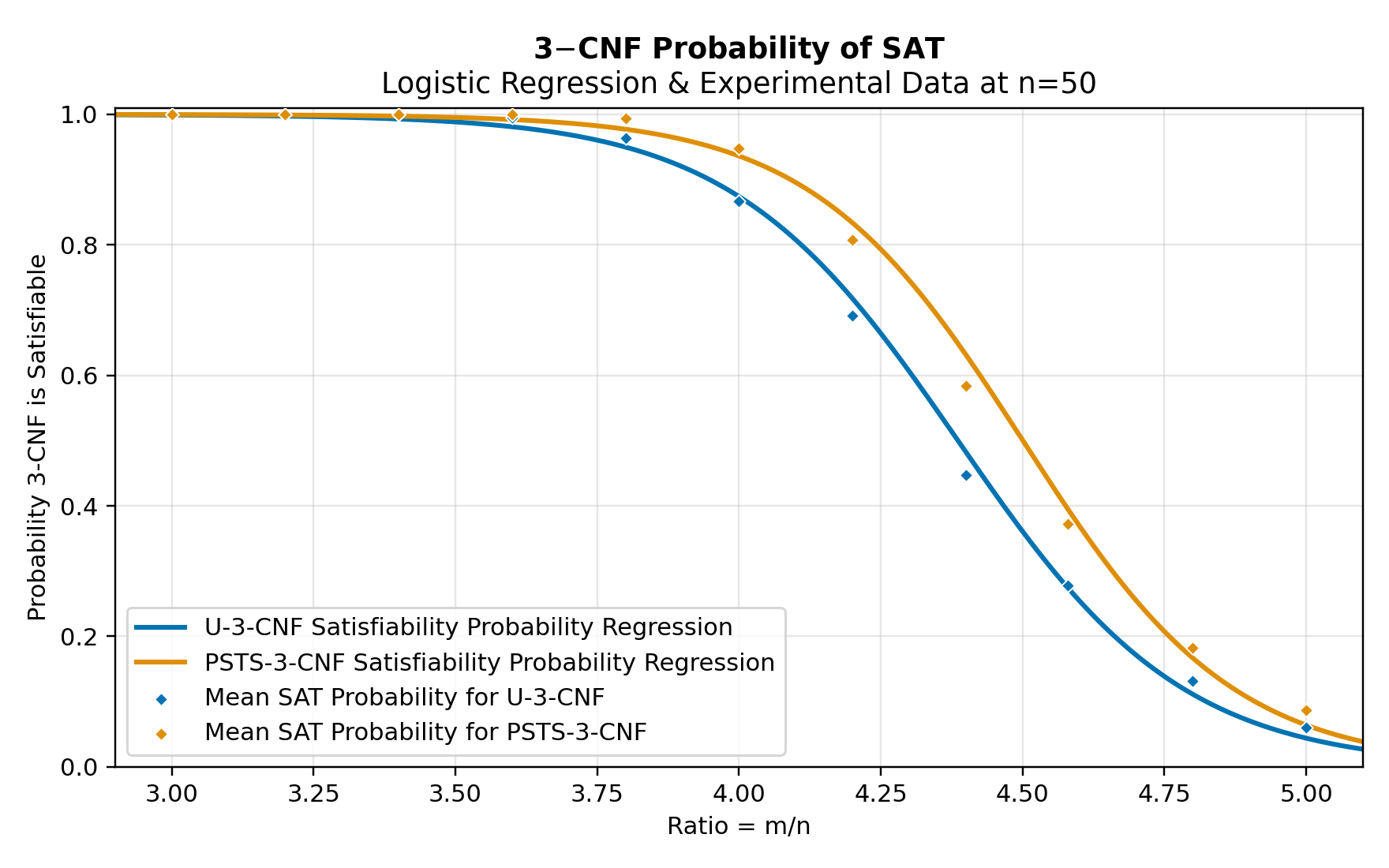}
    \caption{A binomial logistic regression of the probability that a 3-CNF formula is SAT for $n = 50$ (alongside experimental data).
    This is typical of the other $n$ values (See Appendix.).}
    \label{fig:sat_reg_50}
\end{figure}

\section{Results}\label{sec:results}
We use statistical analyses to detect and quantify the differences between PSTS-3-CNF and U-3-CNF in both  probability of satisfiability and number of satisfying assignments given satisfiability.
For our independent variables, we consider the number of binary variables $n$, the ratio of clauses to variables $r = \frac{m}{n}$, and an indicator variable $\mathrm{isPSTS} \in \{0,1\}$, where we set $\mathrm{isPSTS} = 1$ when the 3-CNF formula is a PSTS-3-CNF and $\mathrm{isPSTS} = 0$ otherwise.
We also center our data at $n_0 = 50$, the median $n$ value used in our experiment, and $r_0 = 4.4$, the ratio that corresponds to approximately 50\% satisfiability.
These values of $n_0$ and $r_0$ serve as the center points of our experiment when we ask if PSTS-3-CNF and U-3-CNF are statistically different in both probability of satisfiability and number of satisfying assignments given satisfiability.
We then use $n_c = n - n_0$ and $r_c = r - r_0$ to center $n$ and $r$ respectively.

\subsection{Satisfiability} \label{subsec:satisfiability}

Our first goal is to predict the probability of satisfiability given $n$, $r$, and $\mathrm{isPSTS}$: $\Pr(\mathrm{SAT}\mid n,r,\mathrm{isPSTS})$.
We choose a logistic regression, which is useful when there is a binary outcome (\textit{e.g.}, SAT or UNSAT) and matches the shape of the experimental data (see Figure \ref{fig:sat_reg_50}).
This regression allows us to calculate coefficients $b^{(1)}_1, \cdots, b^{(1)}_7$ for
discrete variables $b^{(1)}_1 n_c$, $b^{(1)}_2 r_c$;
binary variable $b^{(1)}_3 (\mathrm{isPSTS})$;
and their interactions $b^{(1)}_4 (n_c r_c)$, $b^{(1)}_5 (n_c \, \mathrm{isPSTS})$, $b^{(1)}_6 (r_c \, \mathrm{isPSTS})$, and $b^{(1)}_7 (n_c r_c \, \mathrm{isPSTS)}$.
In particular, the coefficient $b^{(1)}_3$ allows us to quantify the effect of $\mathrm{isPSTS}$ on satisfiability: the magnitude of $b^{(1)}_3$ tells us the strength of this effect, the sign of $b^{(1)}_3$ indicates whether PSTS-3-CNF are more or less satisfiable, and the error values associated with $b^{(1)}_3$ show whether $b^{(1)}_3$ itself is statistically significant.
Similarly, $b^{(1)}_5$, $b^{(1)}_6$, and $b^{(1)}_7$ give us this same information about the respective interactions of $\mathrm{isPSTS}$ with $n_c$ and $r_c$.
We use the following logistic regression:

\begin{equation}
\begin{split}
  \log\!\left(\frac{\Pr(\mathrm{SAT}\mid n,r,\mathrm{isPSTS})}{1-\Pr(\mathrm{SAT}\mid n,r,\mathrm{isPSTS})}\right) = b^{(1)}_0 + b^{(1)}_1 n_c + b^{(1)}_2 r_c + b^{(1)}_3 \, \mathrm{isPSTS} + b^{(1)}_4 (n_c r_c) \\ + b^{(1)}_5 (n_c \, \mathrm{isPSTS}) + b^{(1)}_6 (r_c \, \mathrm{isPSTS}) + b^{(1)}_7 (n_c r_c \, \mathrm{isPSTS})
\end{split}
\end{equation}

Table~\ref{tab:logit-sat-results} in the Appendix reports coefficient estimates for the model.
Figure~\ref{fig:sat_reg_50} demonstrates the results of the regression at $n = 50$, which is typical for other values of $n$.\footnote{Other figures can be found in the Appendix.}
As expected, the ratio $r_c$ has a strongly-negative, statistically-significant effect, $\hat{b}^{(1)}_2=-5.0193$, $p<0.001$, on the probability that a 3-CNF formula will be SAT regardless of whether the 3-CNF formula comes from the PSTS or uniform-random distribution.
However, $\mathrm{isPSTS}$ has a positive effect, $\hat{b}^{(1)}_3=0.6094$, $p<0.001$, indicating that PSTS-3-CNF are more likely to be solvable than U-3-CNF.
The $\mathrm{isPSTS}$ effect is weaker than that of the ratio but is not negligible, $|\hat{b}^{(1)}_3=0.6094| < |\hat{b}^{(1)}_2=-5.0193|$.
Furthermore, the interaction of $r_c:\mathrm{isPSTS}$ has a negative, statistically significant effect, $\hat{b}^{(1)}_6=-0.3501$, $p<0.001$.

The other independent variable, $n_c$, and one other interaction of independent variables, $n_c:r_c$, did effect the probability of satisfiability with statistical significance, $p<0.001$.
However, the magnitudes of their coefficients were much smaller, $\hat{b}^{(1)}_1=-0.01725$ and $\hat{b}^{(1)}_4=-0.06451$ respectively.
The remaining interactions $n_c:\mathrm{isPSTS}$ and $n_c:r_c:\mathrm{isPSTS}$ were not statistically significant. This can be seen, perhaps, in the fact that the shapes of the graphs look very similar regardless of the value of $n$ (see Appendix).
Furthermore, this gives some evidence that the statistical significance of PSTS-3-CNF on the probability of satisfiability may hold for larger values of $n$ than this study examined, precisely because all interactions of $n_c$ with $\mathrm{isPSTS}$ were shown to be insignificant.

\begin{figure}
    \centering
     \includegraphics[width=\textwidth]{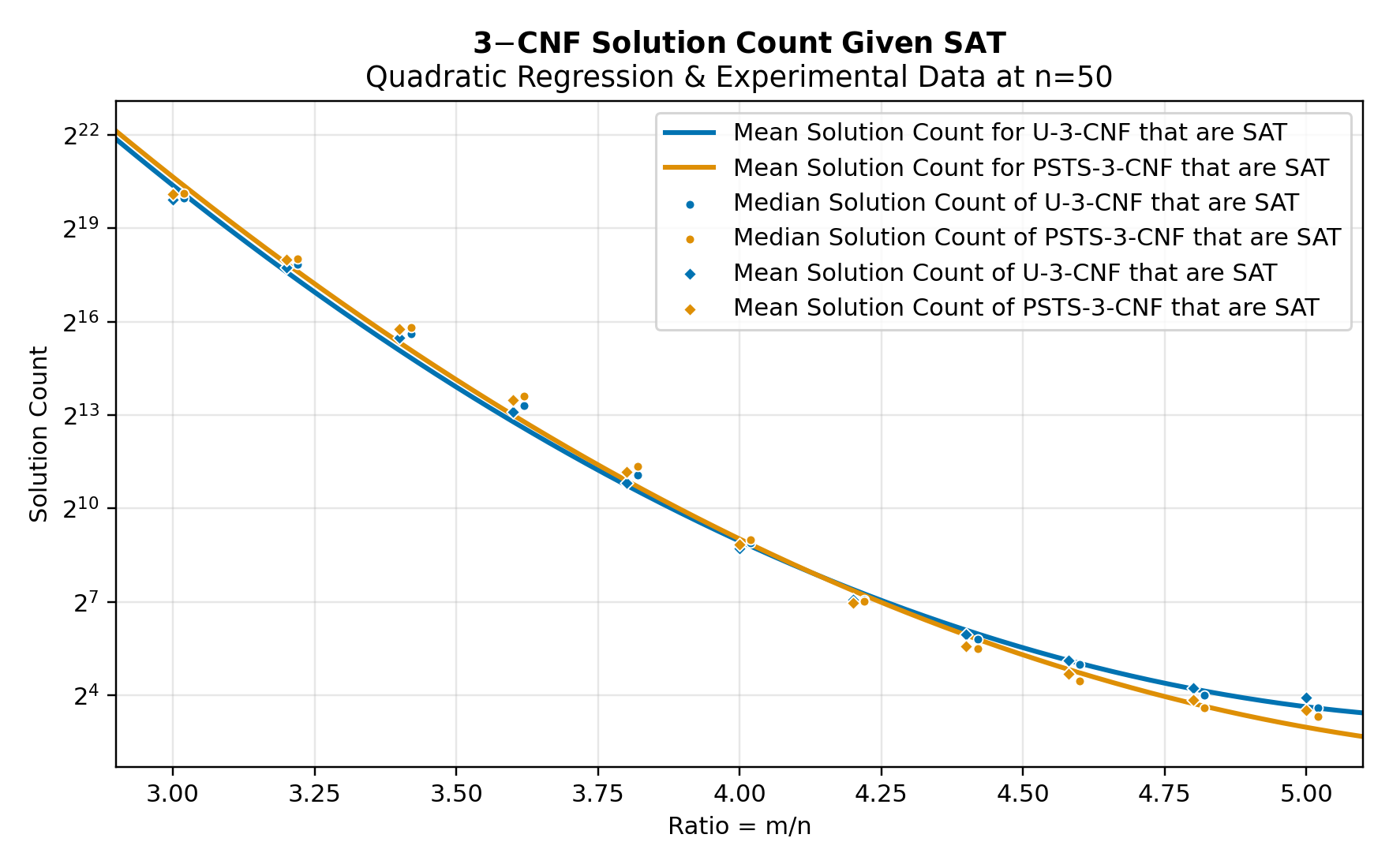}
    \caption{Expected number of satisfying assignments given the 3-CNF formula is SAT for $n = 50$.
    Note that median experimental data (circle icons) are offset slightly on the x-axis to improve visibility.}
    \label{fig:count_reg_50}
\end{figure}

\subsection{Solution Count Given Satisfiability} \label{subsec:count}

A natural follow-up question is if PSTS-3-CNF differ from U-3-CNF in the expected number of satisfying assignments, $\mathbb{E}\!\left[\#\mathrm{sol}\right]$.
To make this question meaningful, we condition our question on satisfiability (i.e., $\mathbb{E}\!\left[\#\mathrm{sol}\mid \mathrm{SAT}\right]$), thus excluding UNSAT 3-CNF.

The number of possible satisfying assignments for a 3-CNF formula grows exponentially in $n$; therefore, we target our regression to predict $\log_2(\#\mathrm{sol})$.
When our experimental data is plotted on a $\log_2(\#\mathrm{sol})$ scale (see Figure \ref{fig:count_reg_50}), the results exhibit positive concavity, so we apply a regression that is quadratic in ratio $r_c$ to capture this curvature.
Similarly to before, we use coefficients $b^{(2)}_1, b^{(2)}_2, \cdots, b^{(2)}_{11}$ to determine how $n_c$, $r_c$, and $\mathrm{isPSTS}$ affect the expected number of satisfying assignments.
We now include terms for $r_c^2$, but exclude the interaction of $r_c$ with $r_c^2$.

\begin{equation}
\begin{split}
  \mathbb{E}\!\left[\,\log_2(\#\mathrm{sol}) \mid \mathrm{SAT}\,\right]
  &= b^{(2)}_0 + b^{(2)}_1 n_c + b^{(2)}_2 r_c + b^{(2)}_3 r_c^2 + b^{(2)}_4 \, \mathrm{isPSTS}
   + b^{(2)}_5 (n_c r_c)
   + b^{(2)}_6 (n_c r_c^2) \\
   &\quad + b^{(2)}_7 (n_c \, \mathrm{isPSTS})
   + b^{(2)}_8 (r_c \, \mathrm{isPSTS})
   + b^{(2)}_9 (r_c^2 \, \mathrm{isPSTS}) \\
   &\quad + b^{(2)}_{10} (n_c r_c \, \mathrm{isPSTS})
   + b^{(2)}_{11} (n_c r_c^2 \, \mathrm{isPSTS})
\end{split}
\end{equation}

We use a weighted least squares regression, giving each 3-CNF a weight inversely proportional to the number of 3-CNF in that bin.
For example, when $n = 50$ (see Figures~\ref{fig:sat_reg_50} and \ref{fig:count_reg_50}) and the ratio is the lowest, $r = 3.0$, there are 10000 SAT PSTS-3-CNF and U-3-CNF each; however, there are only 872 and 599, respectively, when the ratio is the highest, $r = 5.0$.
For the regression, the observations in each of these four bins would be weighted $\frac{1}{10000}, \frac{1}{10000}, \frac{1}{872}$, and $\frac{1}{599}$, respectively.
Otherwise, the more satisfiable ratios would dominate the regression.

Table~\ref{tab:count} in the Appendix contains the coefficient estimates for the model, and Figure~\ref{fig:count_reg_50} graphs the regression at $n = 50$.
The shape of the regression at $n = 50$ is typical for other values of $n$; however, the scale of the y-axis varies significantly with other values of $n$ (See Appendix for graphs of other $n$ values.).

All independent variables and their interactions (except $n_c\!:\!\mathrm{isPSTS}$) were significant with $p < 0.001$, but not all were substantial in magnitude.
As expected, the predicted number of satisfying assignments is dominated by the ratio, $r_c$ ($\hat{b}^{(2)}_2 = -5.9337$) and $r_c^2$ ($\hat{b}^{(2)}_3 = 3.0546$).
Notably, $\mathrm{isPSTS}$ has an effect, mainly through $r_c\!:\!\mathrm{isPSTS}$ ($\hat{b}^{(2)}_8 = -0.6753$), $r_c^2\!:\!\mathrm{isPSTS}$ ($\hat{b}^{(2)}_9 = -0.2637$), and $\mathrm{isPSTS}$ ($\hat{b}^{(2)}_4 = -0.1595$).
Visually, this can be seen in Figure~\ref{fig:count_reg_50} where the expected number of satisfying assignments is higher for PSTS-3-CNF at low ratios, but lower at high ratios.
Although the difference does not appear dramatic, recall that the y-axis is logarithmic.
Finally, it is worth observing that any interaction of $n_c$ with $\mathrm{isPSTS}$ is small or insignificant.

\section{Discussion and Future Work}

Steiner triple systems (STS) and their generalization to partial STS (PSTS) are combinatorial objects that have been studied for generations and continue to be an area of active research, showing promise in a variety of fields outside of mathematics.
Making use of the natural connection between the triples of a (P)STS and the clauses of a 3-CNF formula, we set out to see if 3-CNF formulas derived from (P)STS (PSTS-3-CNF) differ significantly from uniform-random 3-CNF formulas (U-3-CNF) in probability of satisfiability and expected number of satisfying assignments.
Statistical analysis (Section \ref{sec:results}) shows that this is indeed the case, with PSTS-3-CNF more likely to be satisfiable than their U-3-CNF counterparts with an equal number of variables and clauses.
PSTS-3-CNF differ from U-3-CNF in expected solution count as well, but the story is more nuanced.
At lower clause-to-variable ratios, PSTS-3-CNF are likely to have more satisfying assignments than U-3-CNF, but this effect flips at higher ratios.

These results suggest that (P)STS may have something to offer the SAT community, similar to how (P)STS have been employed in other fields.
Although beyond the scope of this paper, (P)STS are often classified based on properties they exhibit or sub-PSTS they contain. 
For example, we might consider how the automorphism group of a (P)STS or the number of Pasch configurations it contains (see Figure \ref{fig:psts_examples}) impacts the satisfiability of its corresponding PSTS-3-CNF. 
Furthermore, some of these properties and configurations could be adapted to apply to 3-CNF formulas more generally.
Our hope is that work along this line will yield fruitful results for the SAT community, such as (P)STS-informed heuristics for solvers.

\section*{Declaration on Generative AI}
The authors used Codex to assist with the programming tasks for experiments and statistical analysis.
After using this tool, the authors reviewed and edited the content as needed and take full responsibility for the publication’s content.

\bibliography{ref}

\appendix
\newpage

\section{Appendix}

\begin{table}[h]
\centering
\caption{A logistic regression of the probability that a 3-CNF is satisfiable.}
\label{tab:logit-sat-results}
\begin{tabular}{lrrrr}
\toprule
Term & Estimate & Std. Error & $z$ & $p$-value \\
\midrule
Intercept      & $-0.0701$ & $0.0039$ & $-18.09$ & $<0.001$ \\
$n_c$          & $-0.01725$ & $0.00039$ & $-44.60$ & $<0.001$ \\
$r_c$          & $-5.0193$ & $0.0119$ & $-420.59$ & $<0.001$ \\
$\mathrm{isPSTS}$            & $\phantom{-}0.6094$ & $0.0056$ & $108.07$ & $<0.001$ \\
$n_c:r_c$      & $-0.06451$ & $0.00117$ & $-55.15$ & $<0.001$ \\
$n_c:\mathrm{isPSTS}$        & $-0.00077$ & $0.00056$ & $-1.37$ & $0.172$ \\
$r_c:\mathrm{isPSTS}$        & $-0.3501$ & $0.0180$ & $-19.46$ & $<0.001$ \\
$n_c:r_c:\mathrm{isPSTS}$    & $\phantom{-}0.00264$ & $0.00177$ & $\phantom{-}1.50$ & $0.135$ \\
\bottomrule
\end{tabular}
\end{table}

\begin{table}[h]
\centering
\caption{A quadratic regression of expected solution count given the 3-CNF is satisfiable.
The dependent variable is $\log_2(\#\mathrm{sol})$.}
\label{tab:count}
\begin{tabular}{lrrrr}
\toprule
Term & Estimate & Robust SE & $z$ & $p$-value \\
\midrule
Intercept                 & $ 6.0902$ & $0.0102$ & $ 595.31$ & $<0.001$ \\
$n_c$                     & $ 0.1097$ & $0.00082$ & $ 134.36$ & $<0.001$ \\
$r_c$                     & $-5.9337$ & $0.0226$ & $-262.55$ & $<0.001$ \\
$r_c^2$                   & $ 3.0546$ & $0.0181$ & $ 168.55$ & $<0.001$ \\
$\mathrm{isPSTS}$                       & $-0.1595$ & $0.0135$ & $ -11.82$ & $<0.001$ \\
$n_c:r_c$                 & $-0.1300$ & $0.00218$ & $ -59.67$ & $<0.001$ \\
$n_c:r_c^2$               & $ 0.0634$ & $0.00173$ & $  36.72$ & $<0.001$ \\
$n_c:\mathrm{isPSTS}$                   & $-0.00037$ & $0.00107$ & $  -0.34$ & $0.732$ \\
$r_c:\mathrm{isPSTS}$                   & $-0.6753$ & $0.0293$ & $ -23.05$ & $<0.001$ \\
$r_c^2:\mathrm{isPSTS}$                 & $-0.2637$ & $0.0236$ & $ -11.17$ & $<0.001$ \\
$n_c:r_c:\mathrm{isPSTS}$               & $-0.0122$ & $0.00281$ & $  -4.34$ & $<0.001$ \\
$n_c:r_c^2:\mathrm{isPSTS}$             & $-0.00890$ & $0.00225$ & $  -3.96$ & $<0.001$ \\
\bottomrule
\end{tabular}
\end{table}

\begin{figure}[p] 
    \centering
    \includegraphics[width=\textwidth]{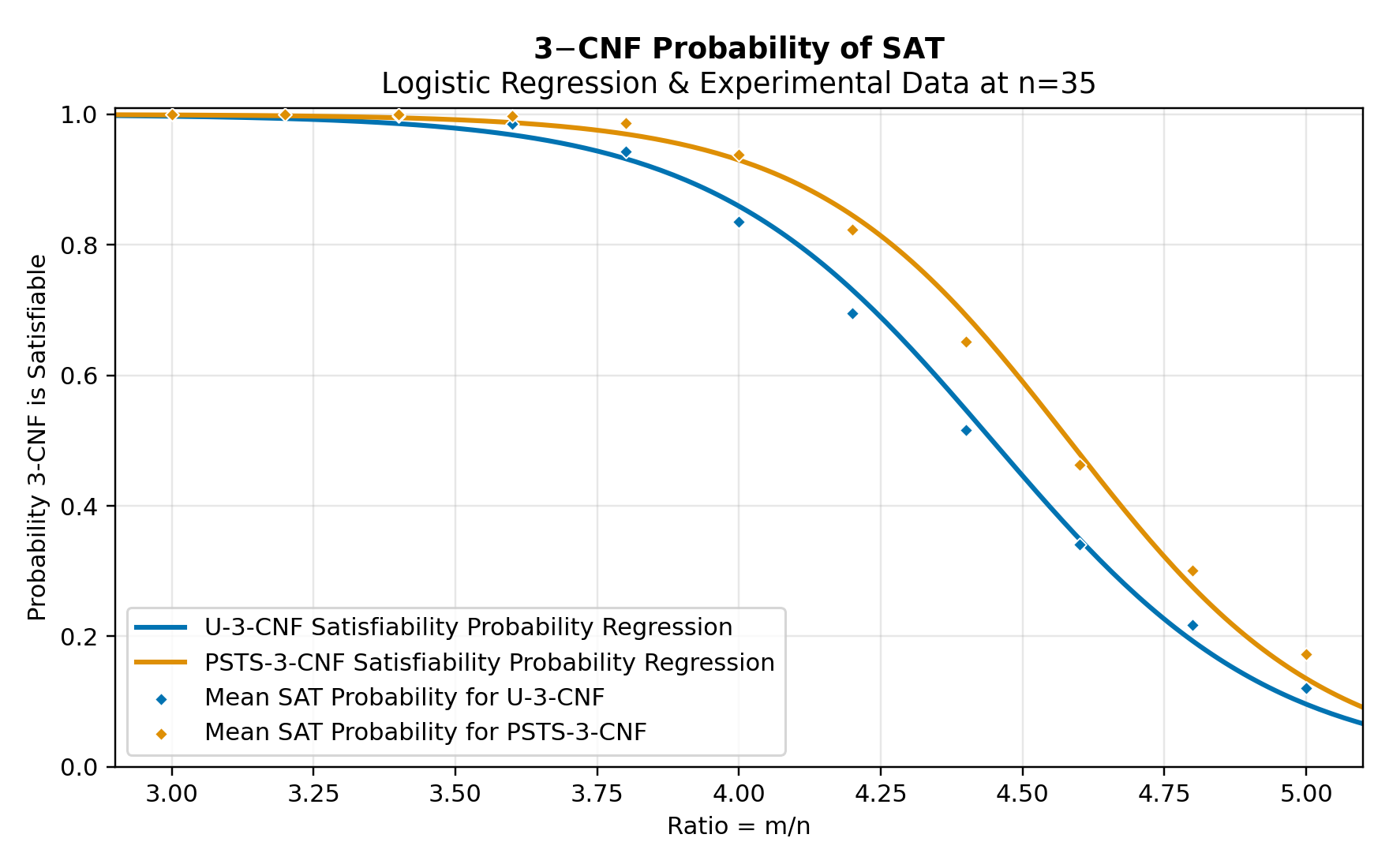}
    \label{fig:apendixA}
    \vspace{1cm}
    \includegraphics[width=\textwidth]{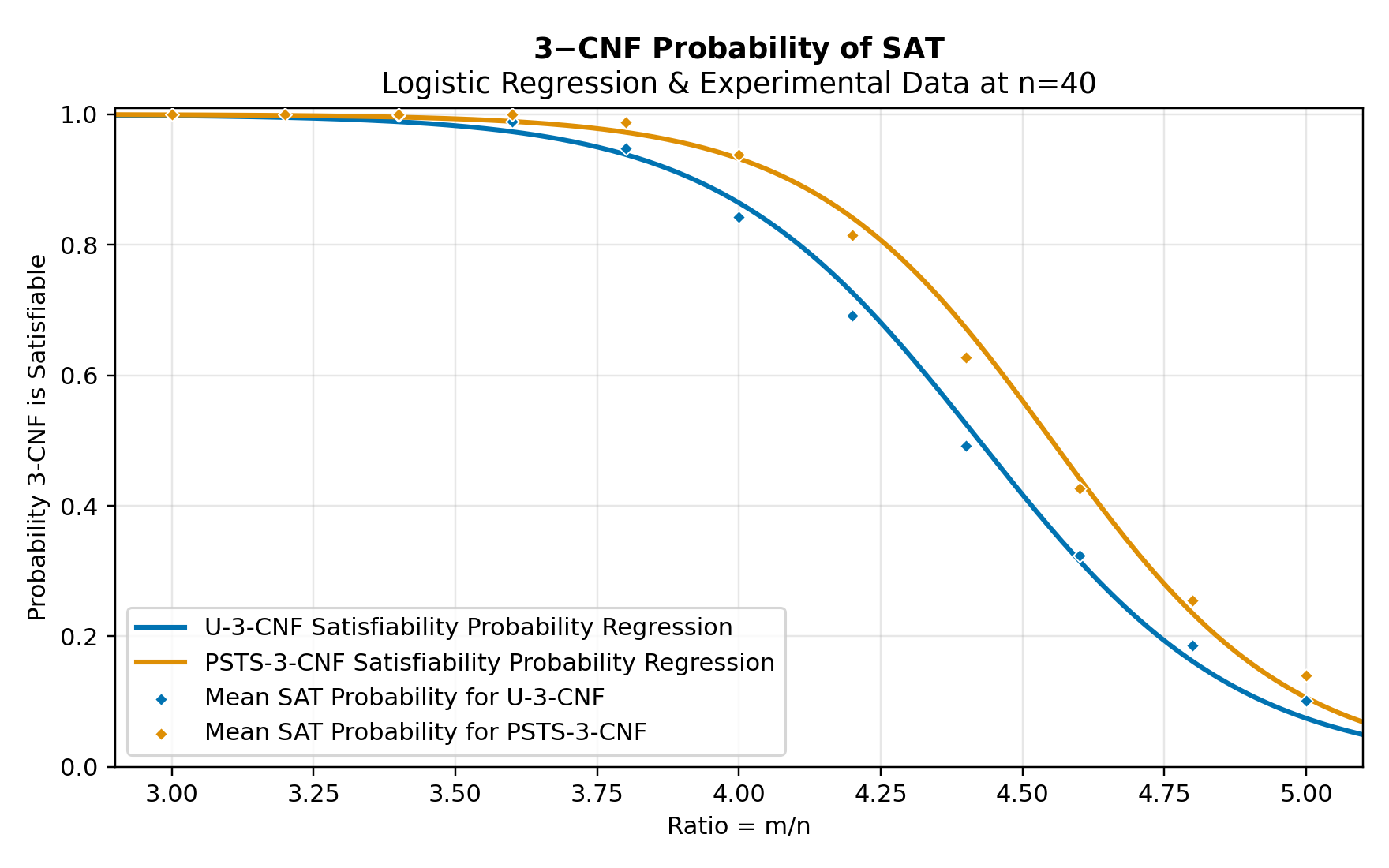}
    \label{fig:appendixB}
\end{figure}

\begin{figure}[p] 
    \centering
    \includegraphics[width=\textwidth]{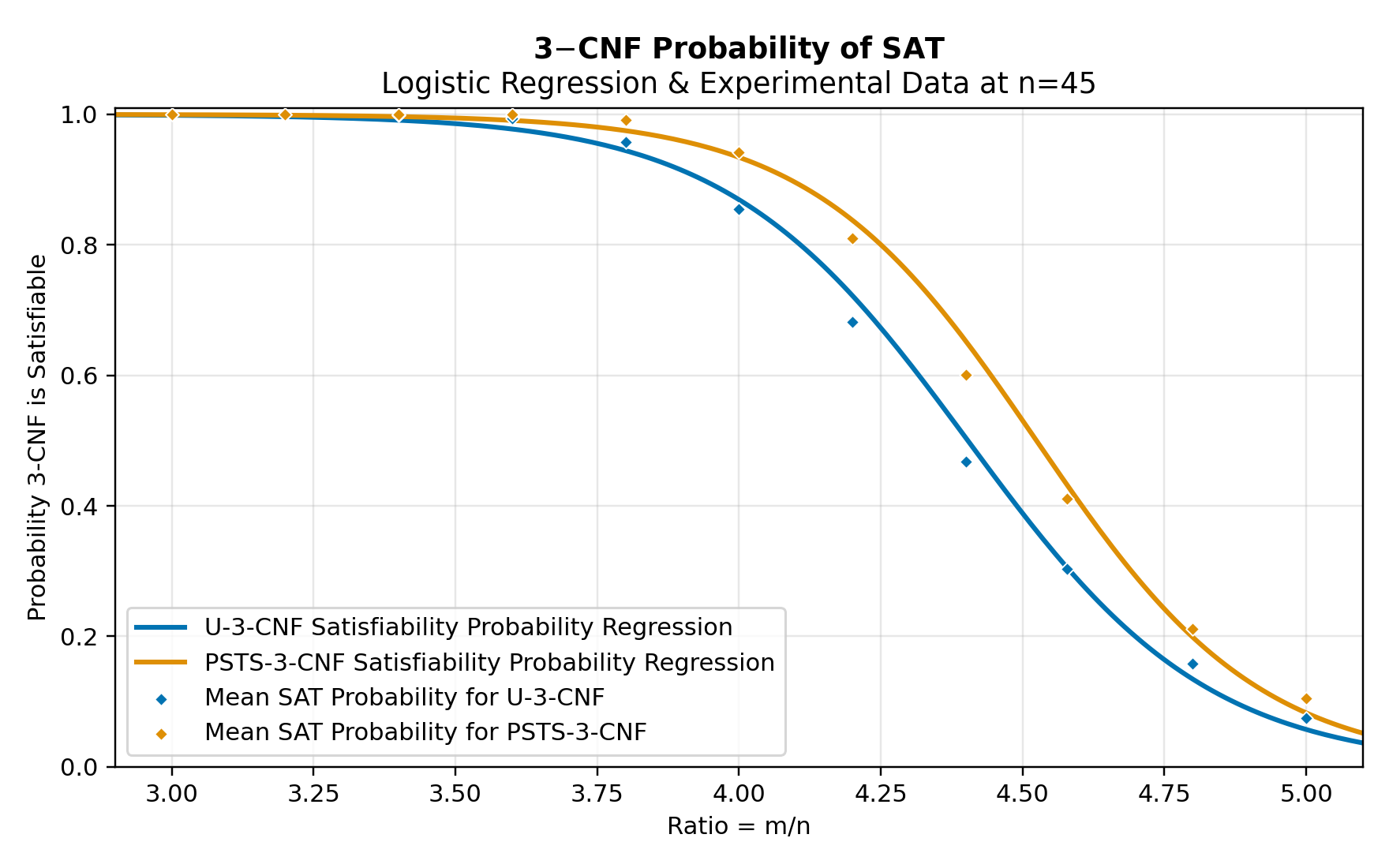}
    \label{fig:apendixC}
    \vspace{1cm}
    \includegraphics[width=\textwidth]{graphs/sat_prop_regression_n50.png}
    \label{fig:appendixD}
\end{figure}

\begin{figure}[p] 
    \centering
    \includegraphics[width=\textwidth]{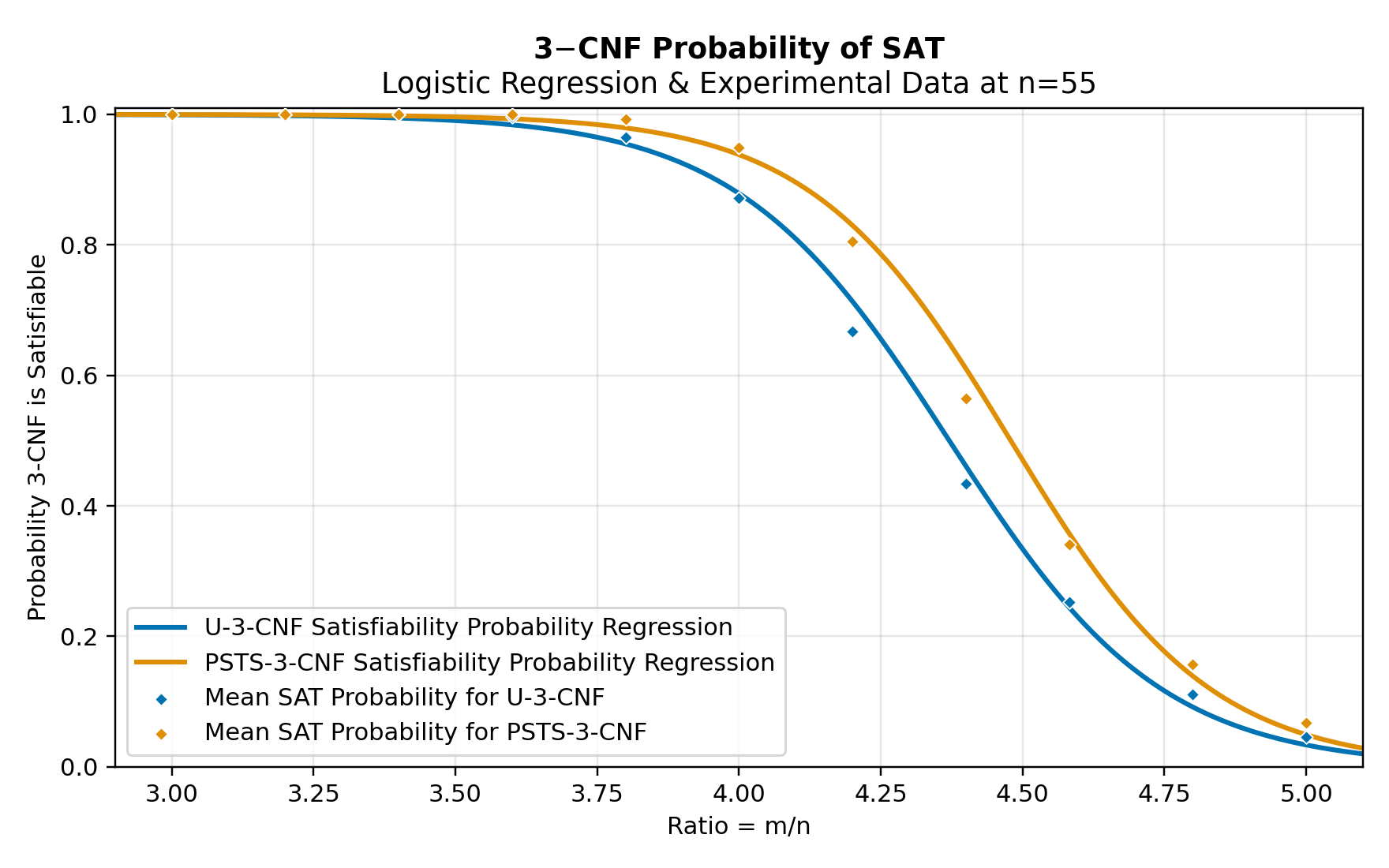}
    \label{fig:apendixE}
    \vspace{1cm}
    \includegraphics[width=\textwidth]{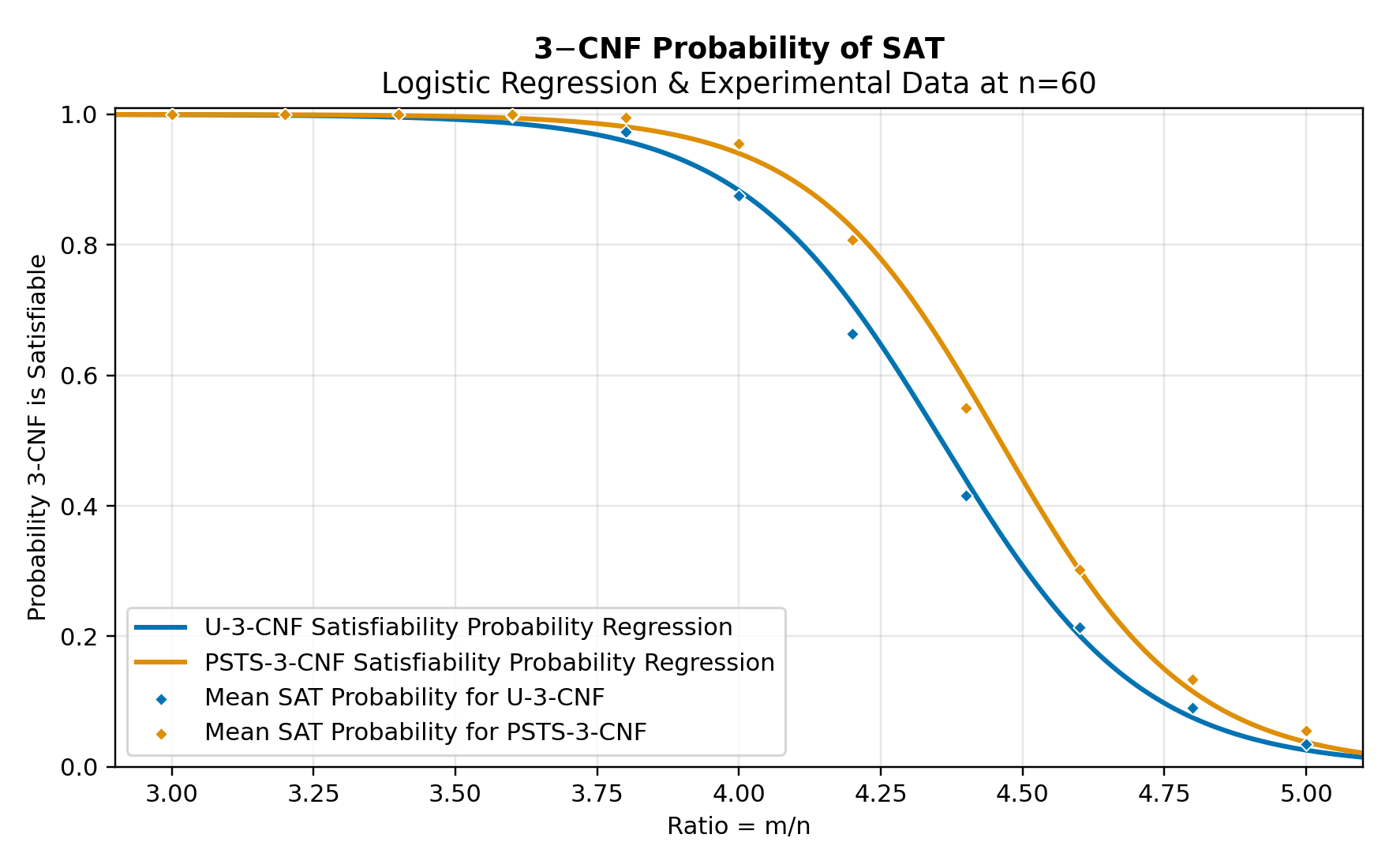}
    \label{fig:appendixF}
\end{figure}

\begin{figure}[p] 
    \centering
    \includegraphics[width=\textwidth]{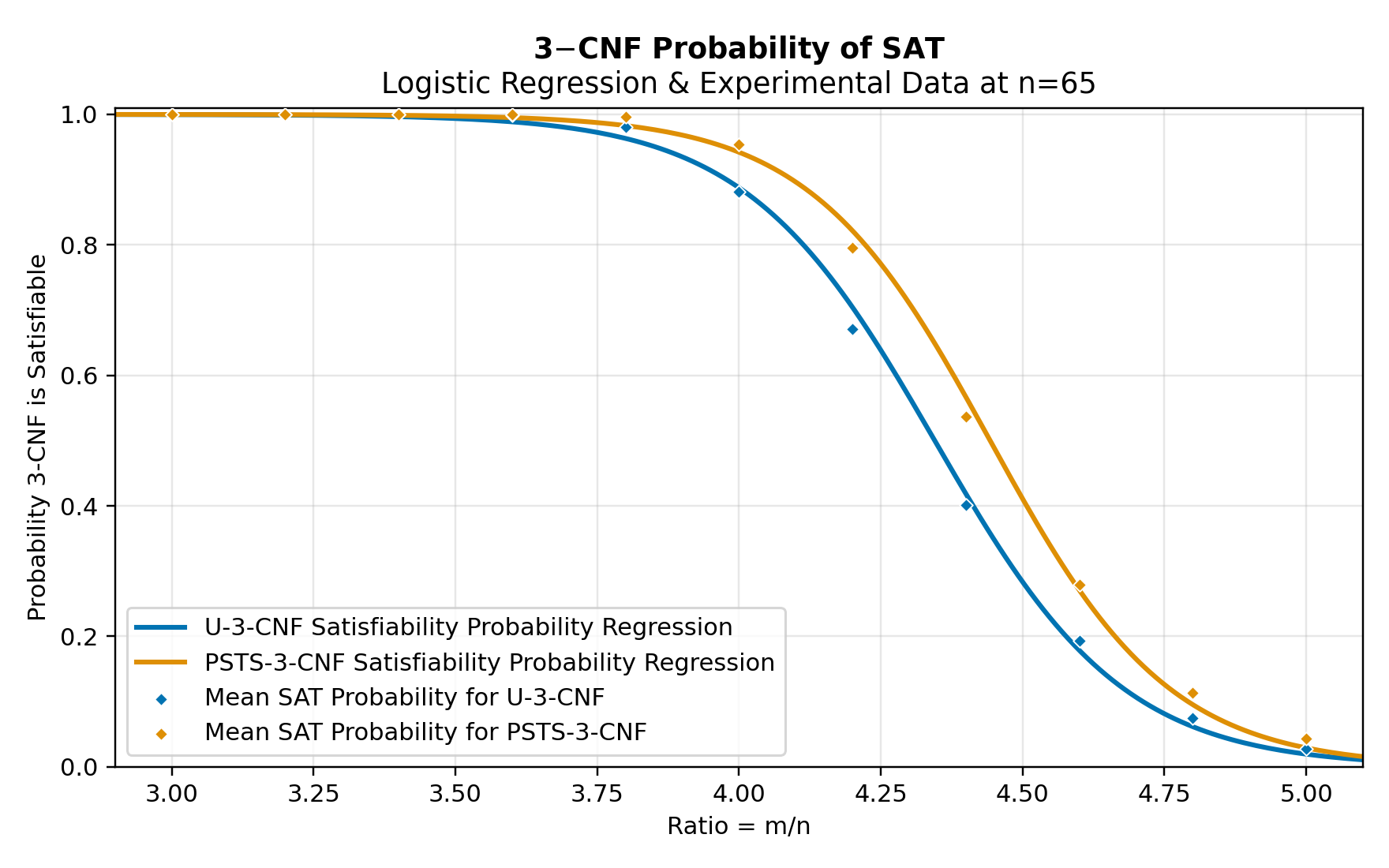}
    \label{fig:apendixG}
    \vspace{1cm}
    \includegraphics[width=\textwidth]{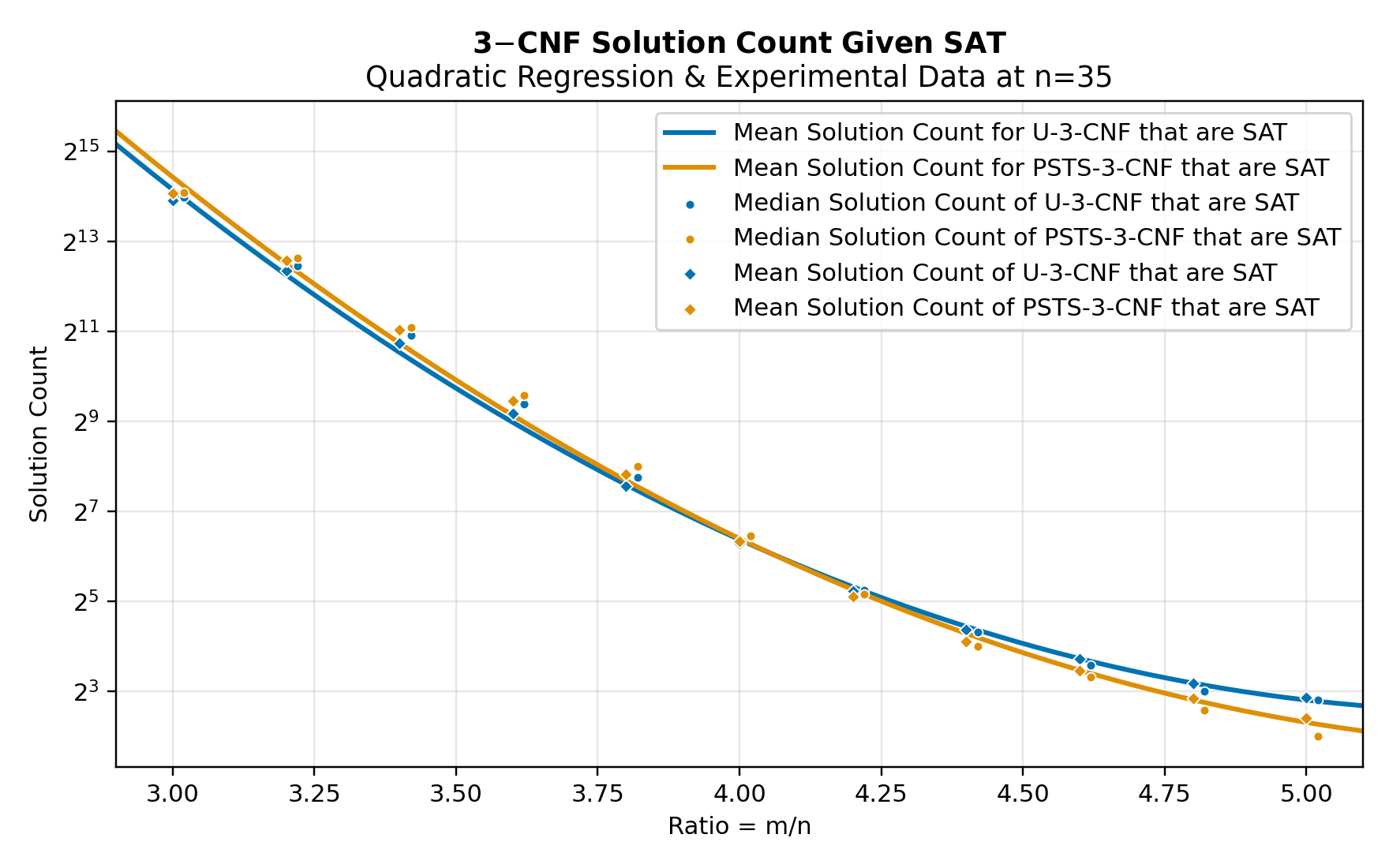}
    \label{fig:appendixH}
\end{figure}

\begin{figure}[p] 
    \centering
    \includegraphics[width=\textwidth]{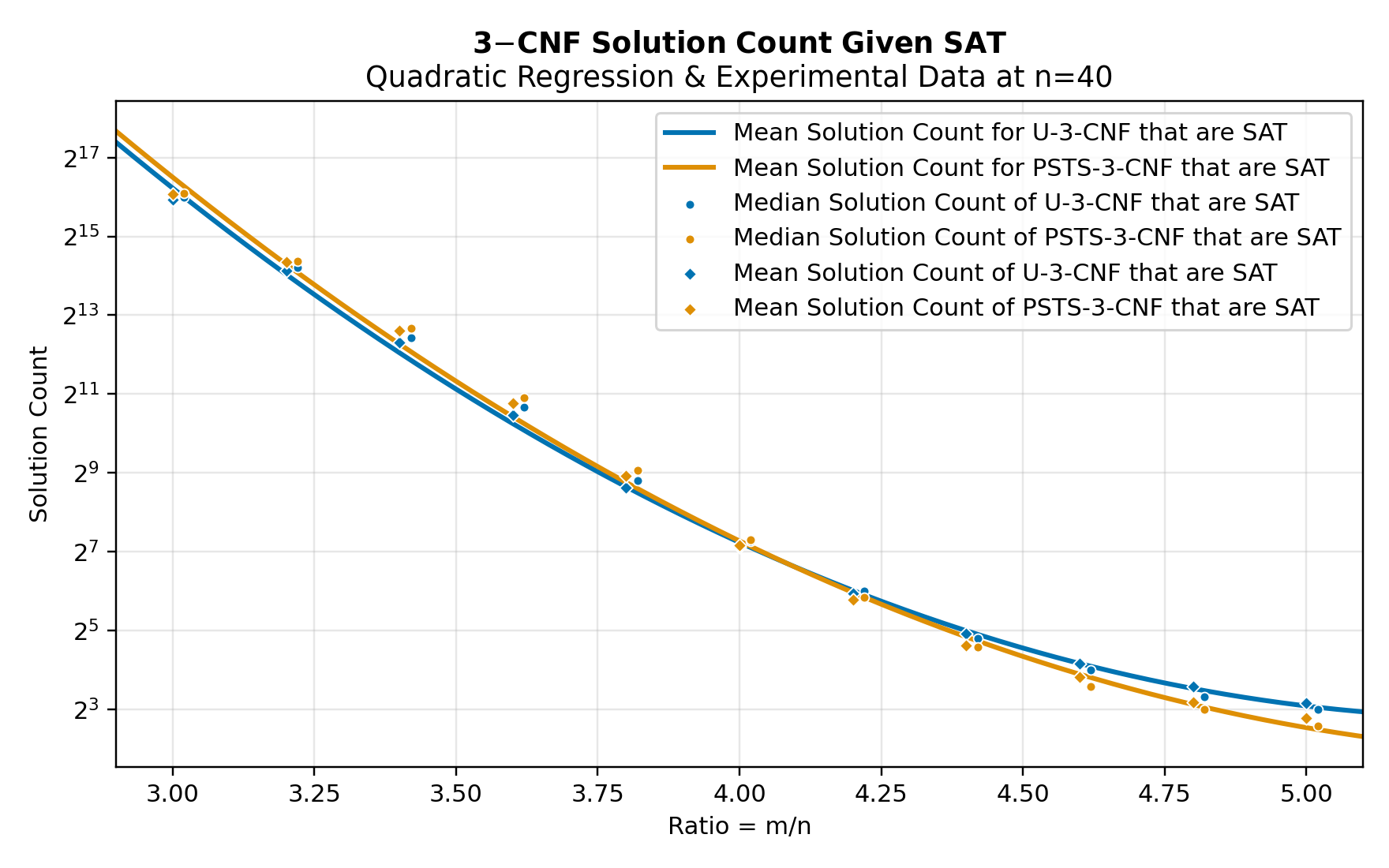}
    \label{fig:apendixI}
    \vspace{1cm}
    \includegraphics[width=\textwidth]{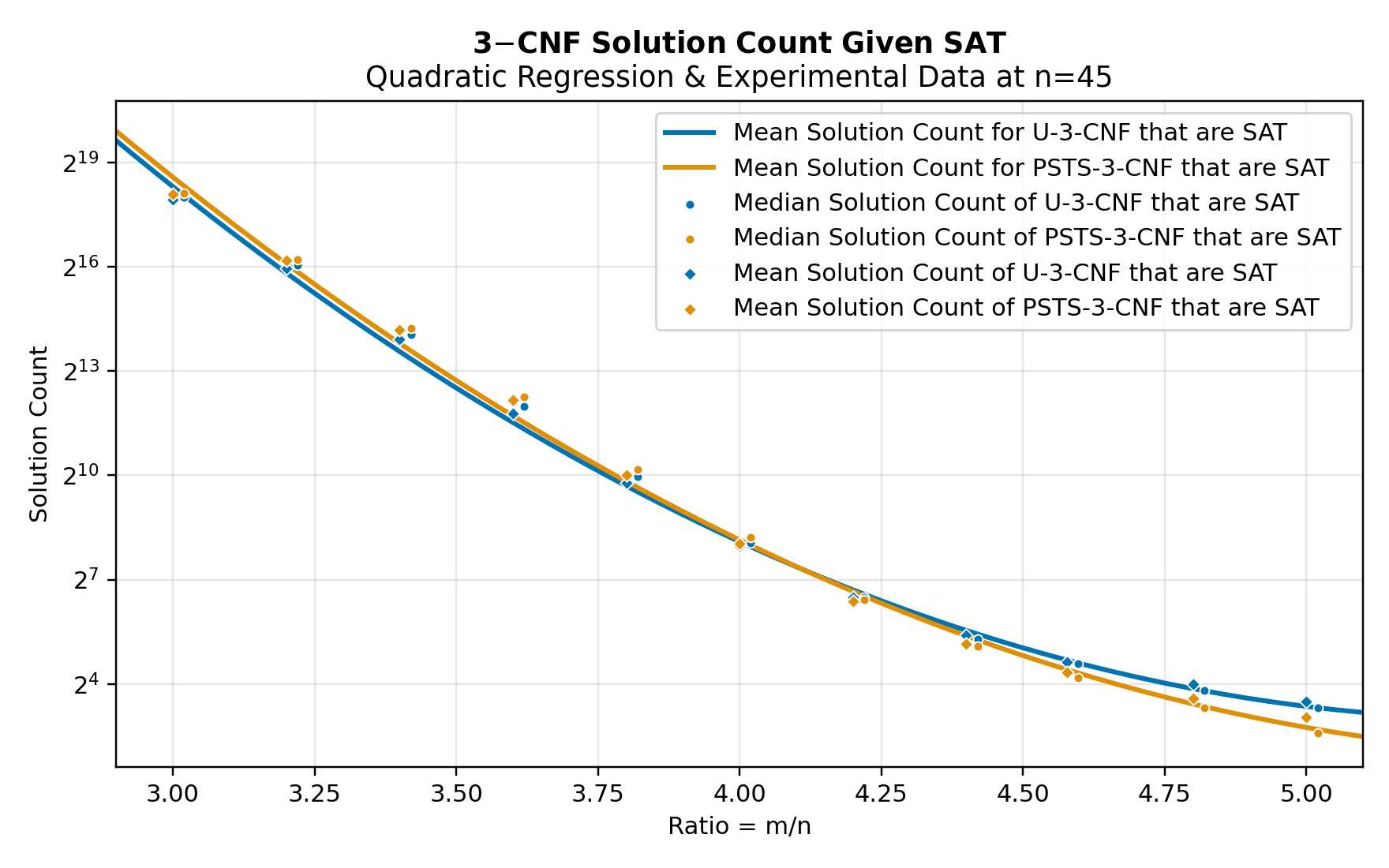}
    \label{fig:appendixJ}
\end{figure}

\begin{figure}[p] 
    \centering
    \includegraphics[width=\textwidth]{graphs/count_regression_n50.png}
    \label{fig:apendixK}
    \vspace{1cm}
    \includegraphics[width=\textwidth]{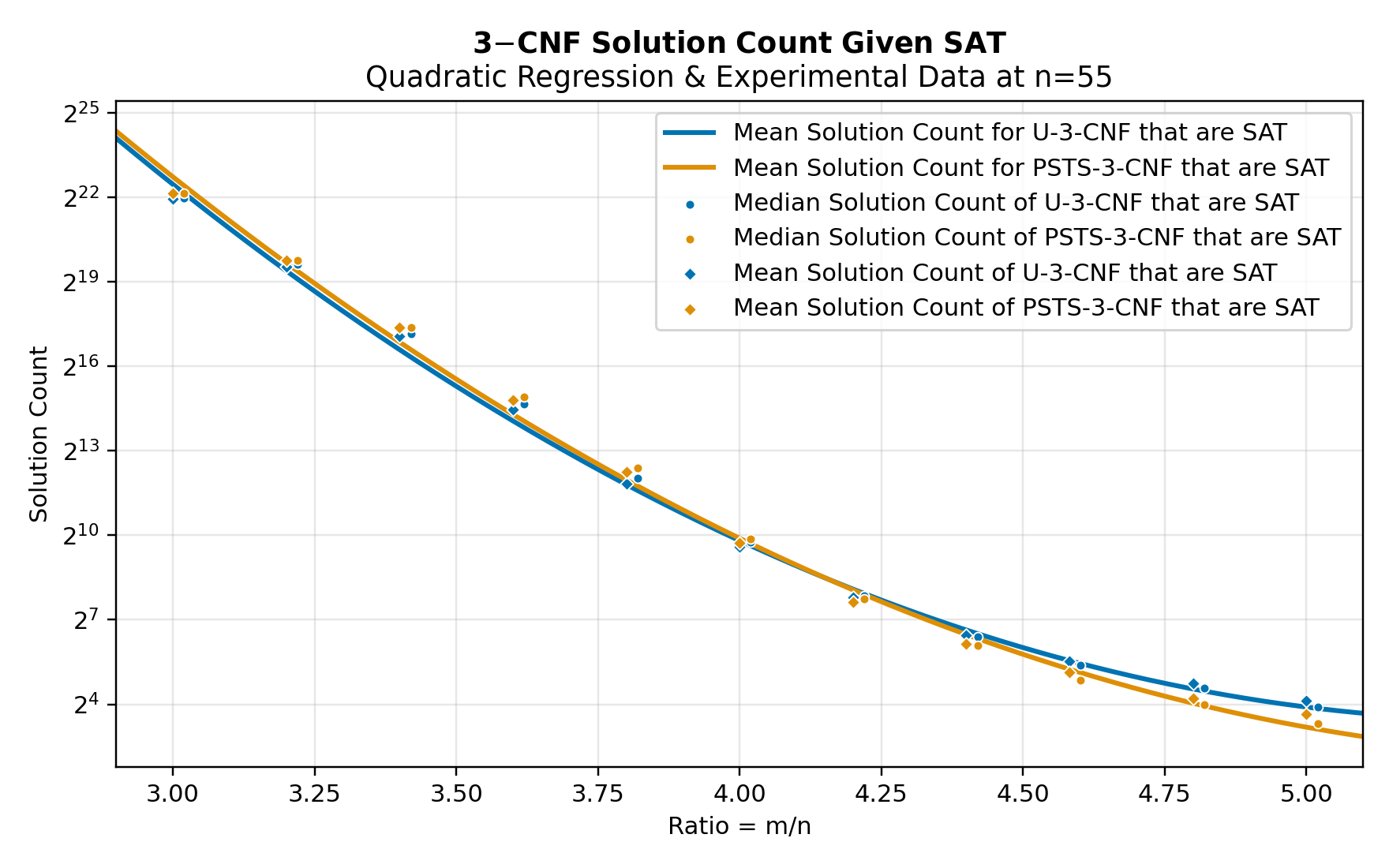}
    \label{fig:appendixL}
\end{figure}

\begin{figure}[p] 
    \centering
    \includegraphics[width=\textwidth]{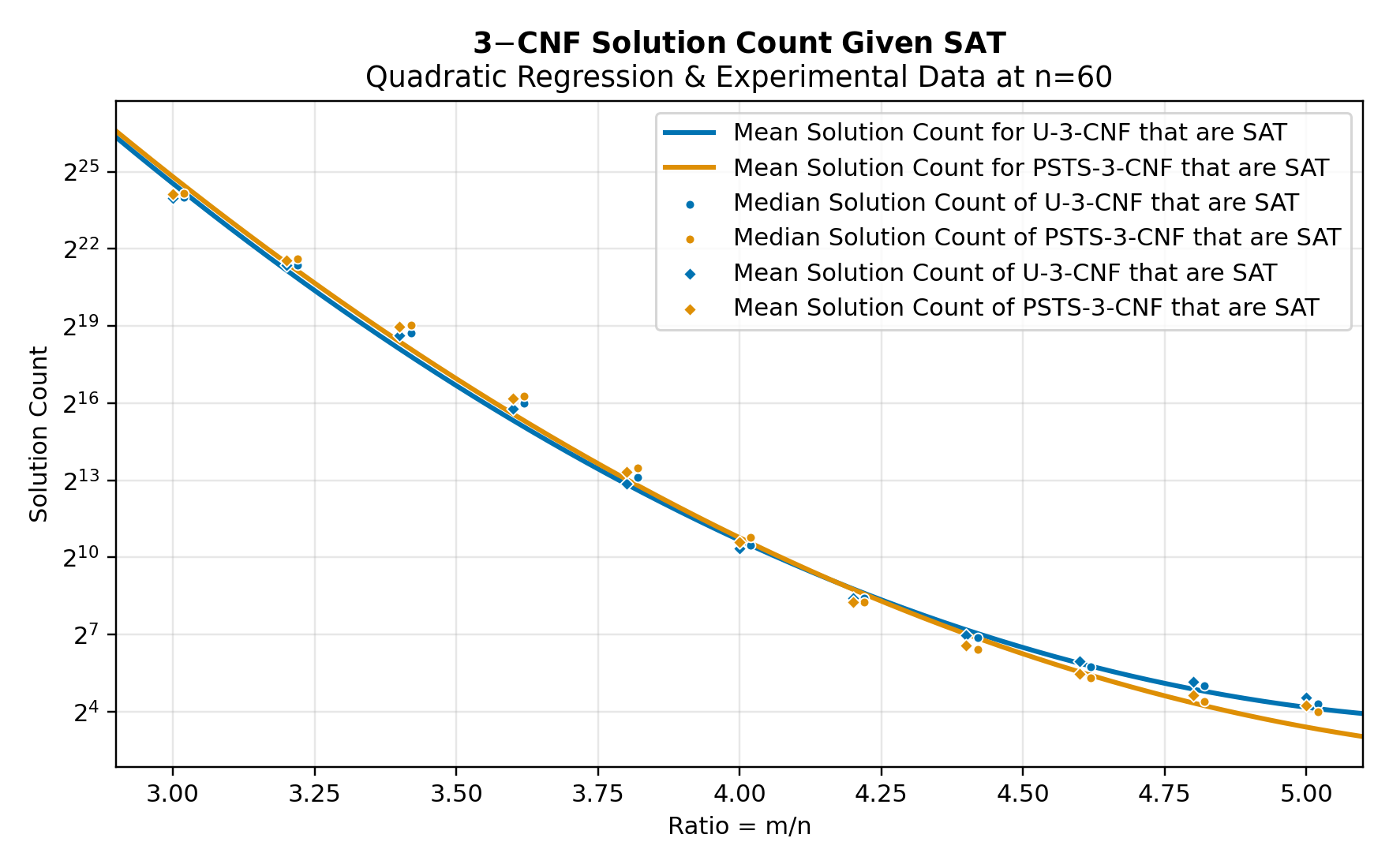}
    \label{fig:apendixM}
    \vspace{1cm}
    \includegraphics[width=\textwidth]{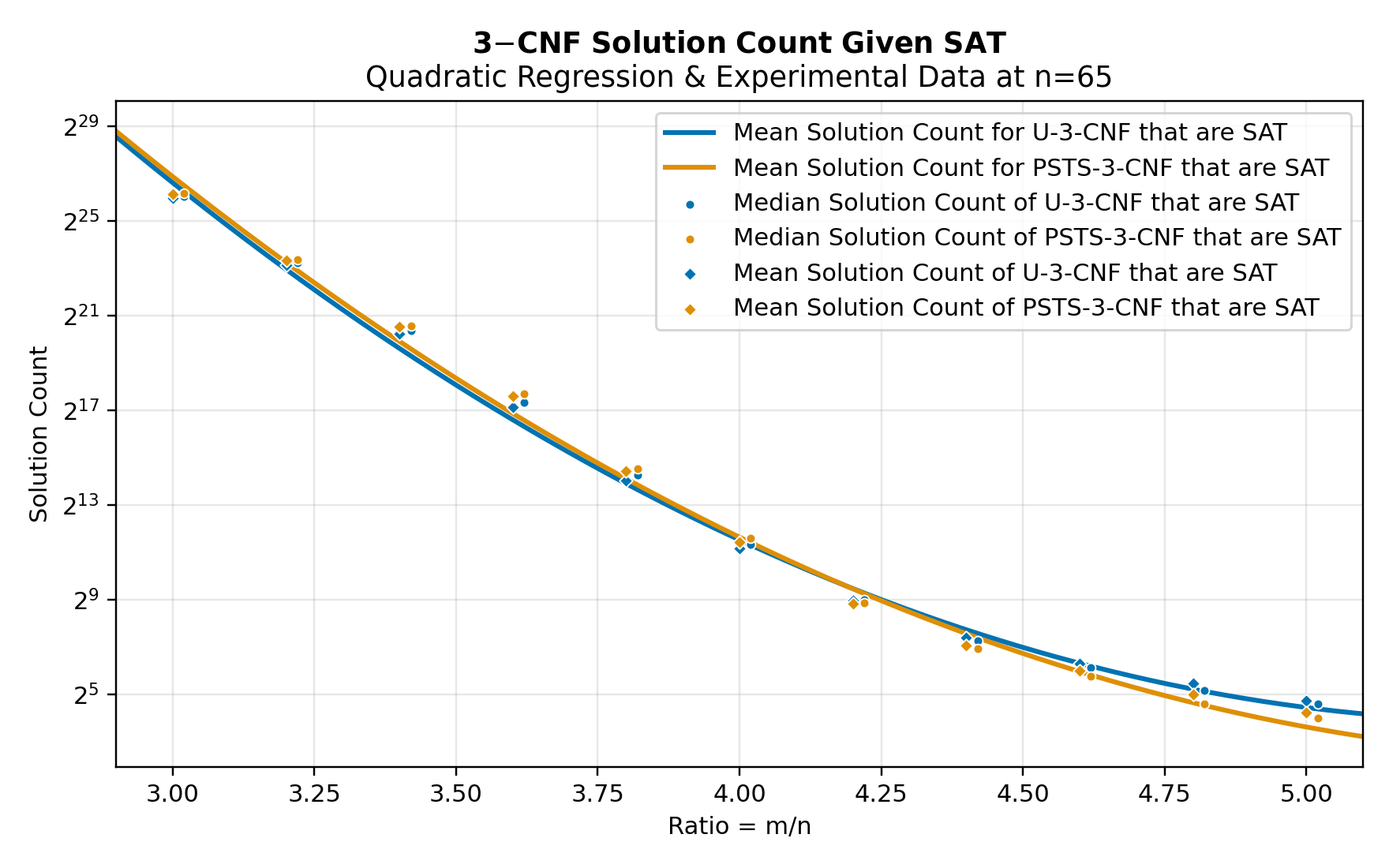}
    \label{fig:appendixN}
\end{figure}

\end{document}